\documentclass[twocolumn,showpacs,preprintnumbers,amsmath,amssymb,floatfix,prd]{revtex4}
\usepackage{epsfig,bm,lscape}
\usepackage{dcolumn}
\usepackage{url}
\begin{document}
\title{Prospects for Charged Current Deep-Inelastic Scattering\\ 
off Polarized Nucleons at a Future Electron-Ion Collider}
\author{Elke C. Aschenauer}\email{elke@bnl.gov}
\affiliation{Physics Department, Brookhaven National Laboratory, Upton, NY~11973, USA}
\author{Thomas Burton}\email{tpb@rcf.rhic.bnl.gov}
\affiliation{Physics Department, Brookhaven National Laboratory, Upton, NY~11973, USA}
\author{Till Martini}\email{martinit@physik.hu-berlin.de}
\affiliation{Institut f\"{u}r Physik, Humboldt-Universit\"{a}t zu Berlin,
D-12489 Berlin, Germany}

\author{Hubert Spiesberger}\email{spiesber@uni-mainz.de}
\affiliation{PRISMA Cluster of Excellence, Institut f\"{u}r Physik, 
Johannes Gutenberg-Universit\"{a}t, Staudingerweg 7, D-55099 Mainz, Germany}
\author{Marco Stratmann}\email{marco@bnl.gov}
\affiliation{Physics Department, Brookhaven National Laboratory, Upton, NY~11973, USA}

\begin{abstract}
We present a detailed phenomenological study of charged-current-mediated
deep-inelastic scattering off longitudinally polarized nucleons at a future
Electron-Ion Collider. 
A new version of the event generator package {\sc Djangoh}, extended by capabilities
to handle processes with polarized nucleons, is introduced and used to
simulate charged current deep-inelastic scattering including QED, QCD, and electroweak radiative effects.
We carefully explore the range of validity and the accuracy of the Jacquet-Blondel method
to reconstruct the relevant kinematic variables from the measured hadronic final state
in charged current events, assuming realistic detector performance parameters.
Finally, we estimate the impact of the simulated charged current single-spin asymmetries on
determinations of helicity parton distributions in the context of a global QCD analysis
at next-to-leading order accuracy.
\end{abstract}

\preprint {HU-EP-13/35, MITP/13-055}
\maketitle

\section{Introduction and Motivation}
%
At sufficiently large momentum transfer $Q^2$, the deep-inelastic scattering
(DIS) process receives significant electroweak contributions where a virtual  
$Z$ or $W^{\pm}$ boson is exchanged between the lepton and the nucleon instead of
a photon. While parity-violating neutral current (NC) reactions are also accessible 
at values of $Q^2$ much smaller than the $Z$ boson mass, thanks to the presence 
of $\gamma Z$-interference contributions to DIS,
charged current (CC) events can only be studied either in high-energy lepton-nucleon 
collisions \cite{Aaron:2009aa} or at neutrino scattering experiments \cite{Tzanov:2009zz}.

Data from CC DIS experiments provide invaluable, complementary information
on the partonic structure of nucleons as they probe combinations
of quark flavors different from those accessible in purely electromagnetic DIS.
In global QCD extractions of unpolarized parton density functions (PDFs) 
CC DIS data help to establish a flavor and 
quark/anti-quark separation \cite{Martin:2009iq,Lai:2010vv,Ball:2012cx}.
Both NC and CC reactions have been studied extensively 
at the DESY-HERA collider using polarized electron and positron beams
scattering off unpolarized protons \cite{Aaron:2009aa,Diaconu:2010zz}. The results confirmed expectations
from electroweak theory by extracting, for instance, up-type and down-type quark couplings.

Corresponding CC DIS data taken on longitudinally polarized nucleons do not yet exist, and 
flavor-separated quark and antiquark helicity PDFs are obtained exclusively from
semi-inclusive DIS data with identified pions and kaons in the final state 
\cite{deFlorian:2008mr,Leader:2010rb}. 
Their theoretical description is more involved than for inclusive DIS and
requires the knowledge of parton-to-hadron fragmentation functions which in turn
have to be extracted from global QCD analyses of inclusive hadron yields \cite{deFlorian:2007aj}.
New data from BNL-RHIC on $W^{\pm}$ production in polarized proton-proton collisions shall
provide an alternative and novel source of information on helicity PDFs at medium-to-large
momentum fractions $x$ \cite{Aschenauer:2013woa}.
Clearly, CC DIS measurements with polarized nucleons 
would be a very welcome and valuable addition to the existing 
suite of experimental data used in extractions of helicity PDFs.

In this paper we perform a detailed study of the feasibility, expected accuracy,
and physics impact of CC DIS measurements on polarized nucleons to be
performed for the first time at a future Electron-Ion Collider (EIC)
\cite{Boer:2011fh,Accardi:2012hwp} such as the eRHIC project at BNL \cite{ref:erhic}.
A major experimental complication and potential limitation compared to purely electromagnetic or
NC DIS is the lack of the scattered lepton to determine the
relevant kinematic variables $x$ and $Q^2$ in CC DIS.
Therefore, we will carefully scrutinize the validity and the accuracy of the Jacquet-Blondel (JB) method
for reconstructing $x$ and $Q^2$ from the measured hadronic final state
in CC events \cite{ref:jb,Blumlein:2012bf}, assuming realistic detector performance parameters.
To simulate polarized CC DIS events, we utilize the event generator package {\sc Djangoh}
\cite{Schuler:1991yg,Charchula:1994kf}, which we have extended to handle 
processes with longitudinally polarized nucleons.
{\sc Djangoh} also allows us to study and quantify the size of
electroweak radiative corrections, in particular QED effects due to
the emission of real photons which can lead to significant shifts of the
kinematic variables away from their ``true'' or Born-level values.
Such radiative corrections are known to be sizable in certain kinematic
regimes from NC and CC DIS measurements at HERA and need to be properly unfolded.

We will demonstrate below that at an EIC one can perform measurements of CC
DIS in the range $x\ \gtrsim 0.02$ $[x\gtrsim 0.01]$ and $Q^2>100\,\mathrm{GeV}^2$,
accessible with the planned lepton and nucleon beam energies of $10\;\mathrm{GeV} \times 250\;\mathrm{GeV}$ 
$[20\;\mathrm{GeV}\times 250\;\mathrm{GeV}]$, with good resolution from the JB method. 
Since the expected CC single-spin asymmetries
are large for most of the accessible $x$ and $Q^2$ region, ranging 
from a few percent at low $x$ up to ${\cal{O}}(80\%)$ at large $x$,
even modest integrated luminosities of ${\cal{L}}=10\,\mathrm{fb}^{-1}$ turn out to be sufficient
for first meaningful measurements. 
We use pseudo-data generated with {\sc Djangoh} in the above kinematic domain
to study their potential impact in constraining helicity PDFs.
To this end, we perform a global QCD analysis at next-to-leading order (NLO) accuracy
following the framework and methodology of the DSSV collaboration \cite{deFlorian:2008mr}.
A similar type of study was performed recently in Ref.~\cite{Aschenauer:2012ve} based on EIC pseudo-data 
for polarized DIS in the low $Q^2$ region dominated by photon exchange.
We note that a first, rough exploratory study of CC DIS at an EIC, 
solely based on simple estimates of expected statistical uncertainties,
has been performed in Sec.~1.12 of Ref.~\cite{Boer:2011fh}.

The remainder of the paper is organized as follows: in the next Section
we shall briefly recall the relevant formalism and expressions for the
CC DIS cross section to define our notation and conventions.
In Sec.~III we introduce the updated event generator package {\sc Djangoh}
which we utilize in Sec.~IV to study the validity and accuracy of the
JB method for reconstructing the relevant DIS kinematic variables 
from the measured hadronic final state.
In Sec.~V we present expectations for the single-spin asymmetries in
CC DIS off polarized protons and neutrons at an EIC and discuss their potential
impact on determinations of helicity parton distributions in the context of a global QCD analysis
at NLO accuracy. The main results are
summarized in Sec.~VI.

\section{Charged Current DIS off Polarized Nucleons \label{sec:theory}}
%
The first theoretical studies of electroweak spin-dependent structure functions date
back to the 1970s, with renewed interest in the HERA era 
\cite{ref:ew-old,deFlorian:1994wp,Stratmann:1995fn} when
the possibility to run with longitudinally polarized proton beams was discussed.
In this context, the first event generator for polarized CC DIS, {\sc Pepsi} 
\cite{Mankiewicz:1991dp,Martin:1997fp},
was developed and some numerical estimates for spin asymmetries at HERA center-of-mass system
(c.m.s.) energies were performed \cite{Maul:1996bs,Contreras:1997fc}, but without including radiative 
effects or scrutinizing the validity of the JB method.
NLO QCD corrections to the polarized CC DIS process have been calculated in \cite{deFlorian:1994wp}.
In this Section we will briefly review the relevant formalism at NLO accuracy
to define the notations and conventions used throughout the paper and otherwise
refer the reader to the PDG review \cite{Beringer:1900zz}.

The spin-dependent part of the CC cross section for the scattering of a left-handed electron ($W^-$ exchange)
off a longitudinally polarized nucleon target $N$ with helicity $\pm \lambda_N$ reads
%
\begin{eqnarray}
\lefteqn{\frac{d^2\Delta\sigma^{W^{-},N}}{dx dy} =} \nonumber \\
&=& \frac{1}{2} \left[ 
\frac{d^2\sigma^{W^{-},N}(\lambda_N=-1)}{dx dy} -
\frac{d^2\sigma^{W^{-},N}(\lambda_N=+1)}{dx dy} \right] 
\nonumber \\
&=&  \frac{2\pi\alpha_{em}^2}{xyQ^2} \eta \Bigg[
2Y_- xg_1^{W^{-},N} - Y_+ g_4^{W^{-},N} + y^2 g_L^{W^{-},N}
\Bigg]
\label{eq:ccxsec}
\end{eqnarray}
where  
\begin{equation}
\eta = 2 
\left( \frac{G_F M_W^2}{4\pi\alpha_{em}} \frac{Q^2}{Q^2 + M_W^2}\right)^2 
\label{eq:propagator}
\end{equation}
and $Y_{\pm}\equiv 1\pm (1-y)^2$.
Here, $M_W$, $G_F$, and $\alpha_{em}$ denote the $W$ boson mass, Fermi constant,
and electromagnetic coupling, respectively, and $Q^2=Sxy$ with $\sqrt{S}$ the available
c.m.s.\ energy.
The corresponding unpolarized CC cross section $d^2\sigma^{W^{-},N}/dxdy$ can be obtained from
(\ref{eq:ccxsec}) by replacing 
$2g_1 \to F_3$, $g_4\to -F_2$, and $g_5 \to -F_1$; see, e.g., Ref.~\cite{Beringer:1900zz}
for details.
We note that Eq.~(\ref{eq:ccxsec}) agrees with the expressions given in
\cite{Beringer:1900zz} except for the extra factor $1/2$ in our definition of
$d^2\Delta \sigma$,
such that the experimentally relevant single-spin asymmetry is defined in the usual way 
as
\begin{equation}
A_L^{W^{-},N} \equiv \frac{d^2 \Delta \sigma^{W^{-},N}/dxdy}{d^2 \sigma^{W^{-},N}/dxdy}
\label{eq:asymdef}
\end{equation}
and will have values $|A_L^{W^{-},N}|\le 1$. 

The structure functions $g_i^{W^{-},N}$ in 
(\ref{eq:ccxsec}) for a proton target and $n_f=4$ active quark flavors are given by
\begin{eqnarray}
\label{eq:g1lo}
g_1^{W^{-},p} (x) &=& \Delta u(x) + \Delta \bar{d}(x) + \Delta c(x)
+\Delta \bar{s}(x) \; ,\\
g_5^{W^{-},p} (x) &=& -\Delta u(x) + \Delta \bar{d}(x) - \Delta c(x)
+\Delta \bar{s}(x)
\label{eq:g5lo} 
\end{eqnarray}
at the leading order (LO) or naive parton model approximation. $g_4$ is related to $g_5$ by the
Dicus relation \cite{Dicus:1972pq}, $g_L\equiv g_4-2x g_5$, with $g_L=0$ at LO (i.e., the
analog to the Callan Gross relation in unpolarized DIS). 
The $\Delta q(x)$ denote the usual helicity parton densities of flavor $q$ in
a longitudinally polarized proton.

The NLO corrections to (\ref{eq:g1lo}), (\ref{eq:g5lo}), and $g_L$ 
can be found in Refs.~\cite{deFlorian:1994wp,Stratmann:1995fn} and can be schematically cast into a simple
form \cite{Forte:2001ph} 
\begin{eqnarray}
g^{{\rm NLO}}_1(x,Q^2)&=&\Delta C_{q,1}\otimes
g^{{\rm LO}}_1+ n_f \,\Delta C_g\otimes\Delta g\;,\nonumber \\[2mm]
\frac{g^{{\rm NLO}}_4(x,Q^2)}{2x}&=&\Delta C_{q,4}\otimes \left[ 
\frac{g^{{\rm LO}}_4}{2x}\right]\; ,\nonumber \\[2mm]
g^{{\rm NLO}}_5(x,Q^2)&=&\Delta C_{q,5}\otimes g^{{\rm LO}}_5\;,
\label{eq:gnlo}
\end{eqnarray}
where the symbol $\otimes$ denotes a convolutional integral which turns into
an ordinary product upon taking Mellin $n$ moments. The latter are
defined as
\begin{equation}
g(n) = \int_0^1 x^{n-1} g(x) dx
\end{equation}
for a function $g(x)$, which is sufficiently regular as $x\to 1$.
The $n$ moments of the relevant coefficient functions $\Delta C_{q,i}$ and $\Delta C_{g,1}$ to NLO accuracy 
in the ${\overline{\mathrm{MS}}}$ scheme are straightforwardly obtained from the $x$
space expressions in \cite{deFlorian:1994wp} and read
\begin{eqnarray}
\nonumber
\Delta C_{q,1} (n) &=& \frac{\alpha_s}{2\pi} C_F \bigg[
                        S_1^2(n) + \left(\frac{3}{2}-\frac{1}{n(n+1)}\right) S_1(n) \\
\nonumber
                         &-&
                         S_2(n) + 
                        \frac{1}{2n} + \frac{1}{n+1} + \frac{1}{n^2} 
                        -\frac{9}{2} \bigg]\\
\nonumber                        
\Delta C_{q,4} (n) &=& \Delta C_{q,1} (n) + \frac{\alpha_s}{2\pi} C_F \left( \frac{1}{n} + \frac{1}{n+1} \right)\\
\nonumber
\Delta C_{q,5} (n) &=& \Delta C_{q,1} (n) + \frac{\alpha_s}{2\pi} C_F \frac{1}{n(n+1)}\\
\Delta C_{g,1} (n) &=& - \frac{\alpha_s}{2\pi} T_F \frac{n-1}{n(n+1)} \left( S_1(n) - \frac{1}{n} +1 \right)
\label{eq:cnlo}
\end{eqnarray}
with $C_F=4/3$, $T_F=1/2$, $S_k(n)=\sum_{j=1}^{n} 1/j^k$,
and $\alpha_s$ the scale-dependent strong coupling; see also \cite{Stratmann:1995fn}.
The $n$ space coefficient functions (\ref{eq:cnlo})
can be straightforwardly implemented into the 
global analysis framework of the DSSV collaboration \cite{deFlorian:2008mr} 
which will be utilized in our phenomenological studies in Sec.~\ref{sec:impact}.

Charged current interactions via $W^+$ exchange probe alternative combinations of
helicity PDFs than in Eqs.~(\ref{eq:g1lo}) and (\ref{eq:g5lo}), 
\begin{eqnarray}
\label{eq:g1lopos}
g_1^{W^{+},p} (x) &=& \Delta \bar{u}(x) + \Delta d(x) + \Delta \bar{c}(x)
+\Delta s(x) \; ,\\
g_5^{W^{+},p} (x) &=& \Delta \bar{u}(x) - \Delta d(x) + \Delta \bar{c}(x)
-\Delta s(x)
\label{eq:g5lopos} 
\end{eqnarray} 
and are only accessible with
positron beams which may or may not be available at a future EIC. 
In lieu of positrons, an effective polarized neutron target in electron DIS, e.g., 
a $^3$He beam with a tag on the spectator protons, also adds valuable, additional
information to a global determination of helicity PDFs.
Assuming, as usual, that the PDFs of the proton and the neutron are related by
$u \leftrightarrow d$ isospin rotation, one probes essentially the same
PDF combinations as in Eqs.~(\ref{eq:g1lopos}) and (\ref{eq:g5lopos})
except for the contributions of the second quark family 
which are sub-leading at the medium-to-large values of $x$ accessible at an EIC.

%
\begin{figure}[h,t]
\begin{center}
\vspace*{-0.6cm}
\epsfig{figure=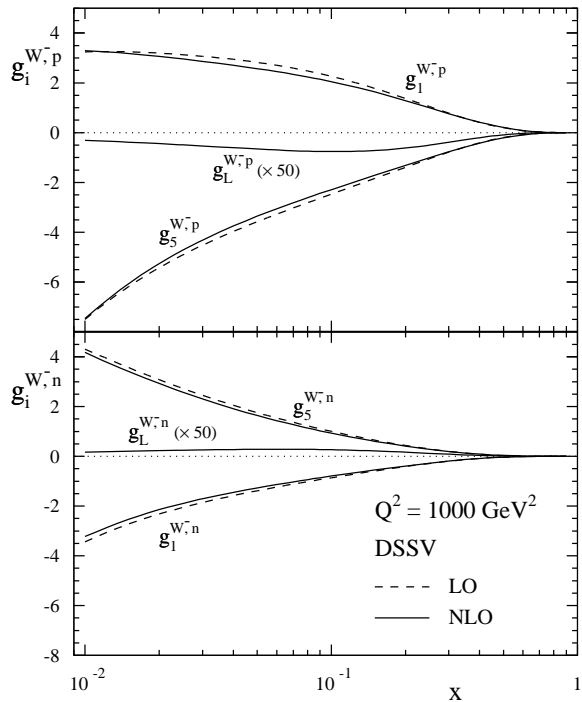,width=0.48\textwidth,angle=0}
\end{center}
\vspace*{-0.6cm}
\caption{\label{fig:gi-nlo} LO and NLO polarized CC DIS structure functions
$g_i^{W^{-},N}$ for protons (top) and neutrons (bottom) as a function of $x$
at $Q^2=1000\,\mathrm{GeV}^2$ using the DSSV helicity PDFs \cite{deFlorian:2008mr}.}
\end{figure}
In Fig.~\ref{fig:gi-nlo} we show the size of the NLO corrections
to the polarized structure functions
$g_i^{W^{-},N}$ for protons (top) and neutrons (bottom) at a typical
value of $Q^2$ and range in $x$ relevant for CC DIS measurements at an EIC.
As can be seen, NLO effects are in general rather modest, and the breaking of the
Dicus relation is numerically very small; note that the curves for $g_L$ are
scaled by a factor of 50 to make them visible. Corresponding QCD corrections
for unpolarized CC structure functions are also small and, as we shall show below
in Sec.~\ref{sec:impact}, almost completely cancel in the experimentally
relevant spin asymmetry $A_L^{W^{-},N}$ defined in Eq.~(\ref{eq:asymdef}).
Also, notice that the sign of the polarized structure functions
$g_i^{W^{-},N}$ flips upon $p\leftrightarrow n$ isospin rotation as
$\Delta u(x)>0$ and $\Delta d(x)<0$ for all sets of helicity PDFs.

Finally, we wish to recall the existence of novel sum rules satisfied by CC 
structure functions, 
which are equally fundamental as the Bjorken sum rule \cite{Bjorken:1966jh} in pure photon exchange.
For instance, one finds, including NLO QCD corrections \cite{Stratmann:1995fn},
\begin{equation}
\int_0^1 dx \left[ g_5^{W^-,n}-g_5^{W^-,p} \right] 
= \left( 1-\frac{2 \alpha_s}{3 \pi} \right) g_A \; ,
\label{eq:sumrule}
\end{equation}
where the superscripts $p$ and $n$ indicate measurements to be taken on proton and
neutron targets, respectively, and $g_A$ represents the axial charge.
Unfortunately, such sum rules are likely of limited phenomenological relevance.
Firstly, there will be a perhaps substantial uncertainty from extrapolating $g_5$
to the small $x$ region in order to evaluate the integral in (\ref{eq:sumrule}).
Secondly, to make use of (\ref{eq:sumrule}) one would need to disentangle the
structure function $g_5$ from the cross section (\ref{eq:ccxsec})
based on the different $y$ dependences in $Y_{\pm}$. Such a ``Rosenbluth separation'' 
requires measurements at fixed $x$ and $Q^2$ but variable $S$ 
which is certainly challenging.

\section{The updated event generator package DJANGOH \label{sec:django}}
%
The event generator package {\sc Djangoh} \cite{Charchula:1994kf}
is an interface to {\sc Heracles} \cite{Kwiatkowski:1990es} for 
the simulation of DIS including electroweak radiative corrections 
with {\sc Lepto} \cite{Ingelman:1996mq}, which implements string 
fragmentation from the {\sc Jetset} library \cite{Sjostrand:1993yb} 
for the simulation of the hadronic final state. First-order QCD 
parton cascades are modeled by the {\sc Ariadne} \cite{Lonnblad:1992tz} 
program. 

Previous versions of {\sc Djangoh}, which were routinely used 
at HERA by the experimental collaborations to correct DIS data 
for electroweak higher-order effects, were restricted to unpolarized 
proton beams. For the present analysis, a new version has been 
developed which allows one to also study the deep-inelastic 
scattering off longitudinally polarized hadron beams. 

The implementation of higher-order corrections at one-loop order 
and including one-photon radiative effects is straightforward 
in the case of CC scattering since for massless 
quarks helicity agrees with chirality. A 
proper replacement of unpolarized PDFs by their polarized counterparts is 
therefore sufficient. {\sc Djangoh} includes a corresponding 
interface to a set of publicly available parametrizations for 
polarized PDFs and provides the required grid files. 

In addition, although not pursued here, 
there are possibilities to simulate scattering off heavy nuclei. 
Nuclear mass number and charge can be chosen 
arbitrarily, and various models for nuclear shadowing can be selected. 
For instance, one option implements a simple $Q^2$-independent shadowing which 
can be imposed on any set of PDFs; other options use specifically 
designed nuclear PDFs as provided by the {\sc Lhapdf} library 
\cite{Whalley:2005nh}. 

For more details we refer to the documentation on the {\sc Djangoh} 
website \cite{djangoh-url}, from where also the code can be obtained.

%
\begin{figure}[ht]
\begin{center}
\vspace*{-0.6cm}
\epsfig{figure=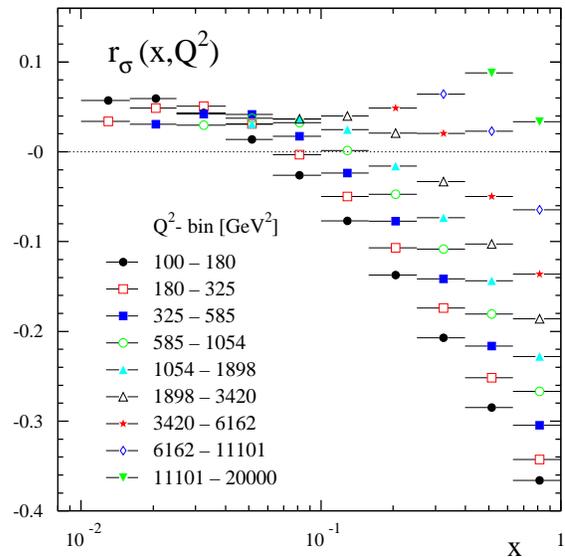,width=0.48\textwidth,angle=0}
\end{center}
\vspace*{-0.6cm}
\caption{\label{fig:rc1} [color online] 
Radiative correction factor $r_\sigma$ as defined in (\ref{eq:rsigma})
for unpolarized CC electron scattering off protons for $x$-$Q^2$ bins 
accessible at an EIC with $\sqrt{S}\simeq141\,\mathrm{GeV}$.}
\end{figure}
\begin{figure}[h]
\begin{center}
\vspace*{-0.6cm}
\epsfig{figure=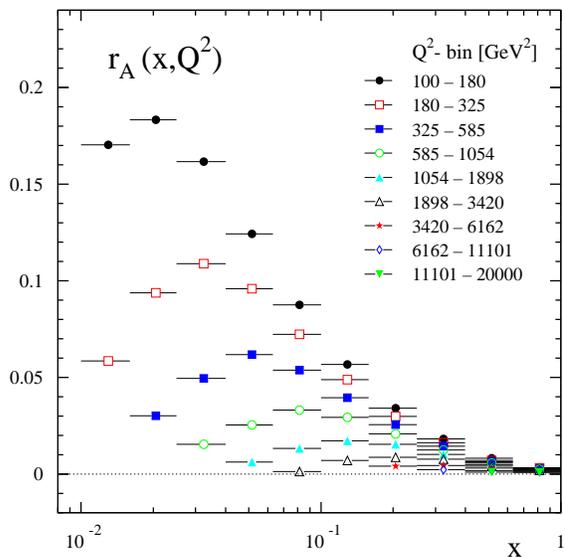,width=0.48\textwidth,angle=0}
\end{center}
\vspace*{-0.6cm}
\caption{\label{fig:rc2} [color online] 
As in Fig.~\ref{fig:rc1} but now for the single-spin asymmetry
$A_L^{W^{-},p}$.}
\end{figure}
As long as one is not interested in a simulation of the hadronic 
final state, QED radiative corrections can be studied with the 
help of the programs {\sc Radgen}, {\sc Polrad} 
\cite{Akushevich:1997di,Akushevich:1998ft,Akushevich:1998dz,Afanasev:2001zg}, 
or {\sc Hector} \cite{Arbuzov:1995id,Bardin:1996ch}.  
Early studies of CC DIS with a polarized proton beam at HERA 
\cite{Maul:1996bs,Contreras:1997fc} were based on the LO
Monte Carlo program {\sc Pepsi} \cite{Mankiewicz:1991dp}. A later 
version of {\sc Pepsi} \cite{Martin:1997fp} included electroweak 
corrections, using {\sc Hector}, but was never used for published 
phenomenological studies. The new version of {\sc Djangoh}
supersedes the {\sc Pepsi} generator. 

The analysis described in this paper is based on an event simulation 
for CC DIS using {\sc Djangoh} including radiative effects. In Fig.~\ref{fig:rc1} 
we show the radiative correction factor for the 
unpolarized CC cross section, 
\begin{equation}
\label{eq:rsigma}
r_\sigma = d^2 \sigma^{W^{-},p}|_{{\cal{O}}(\alpha_{em}^3)} / 
d^2 \sigma^{W^{-},p}|_{{\cal{O}}(\alpha_{em}^2)} - 1,
\end{equation}
for the binning in $x$ and $Q^2$ used in our phenomenological
analysis below, assuming a c.m.s.\ energy of 
$\sqrt{S}\simeq 141\,\mathrm{GeV}$ which corresponds to
lepton and nucleon beam energies of 
$20\;\mathrm{GeV}\times 250\;\mathrm{GeV}$ at the eRHIC
option of an EIC \cite{ref:erhic}.
Here, $x$ and $Q^2$ refer to the leptonic variables 
at the generator level. $r_\sigma(x,Q^2)$ exhibits a behavior 
known from NC scattering: positive corrections at small $x$, i.e., 
at large $y$ for fixed $Q^2$. Since the phase space for photon 
emission is shrinking towards large $x$, one observes large 
negative corrections, dominated by virtual contributions, as 
$x \rightarrow 1$. 
We should emphasize that the actual size of  
radiative effects strongly depends on the prescription used to 
reconstruct kinematic variables. The numerical results shown 
here are meant as an illustration of the possible importance 
of radiative corrections, but will be numerically different when 
they are evaluated within a realistic analysis where kinematic 
variables are reconstructed from the hadronic final state, as 
described in the next Section. 

QED is invariant with respect to parity and the probability to emit a
photon, which can be described by a ``radiator function'', does not depend
on the chirality of the emitting particle. Nevertheless, QED effects do not
cancel completely in the single-spin asymmetry since the corrections
are convolutions of the radiator functions with partonic cross sections
and spin-dependent PDFs. 
Numerical results for 
$r_A = A_L^{W^-,p}|_{{\cal{O}}(\alpha_{em}^3)} / 
A_L^{W^-,p}|_{{\cal{O}}(\alpha_{em}^2)} - 1$
in the same $x$ and $Q^2$ bins as above are given in Fig.~\ref{fig:rc2}. 
The ${\cal{O}}(\alpha_{em})$ corrections differ only by a few percent 
between the two helicity cross sections in Eq.~(\ref{eq:ccxsec})
and are therefore negligible where $A_L^{W^-,N}$ 
is large, but they can become important at smaller $x$
where $A_L^{W^-,N}$ is small at tree-level as we shall see below.

\section{Reconstruction of $x$ and $Q^2$ from the hadronic final state}
%
\begin{figure*}[bht]
\begin{center}
\vspace*{-0.5cm}
\epsfig{figure=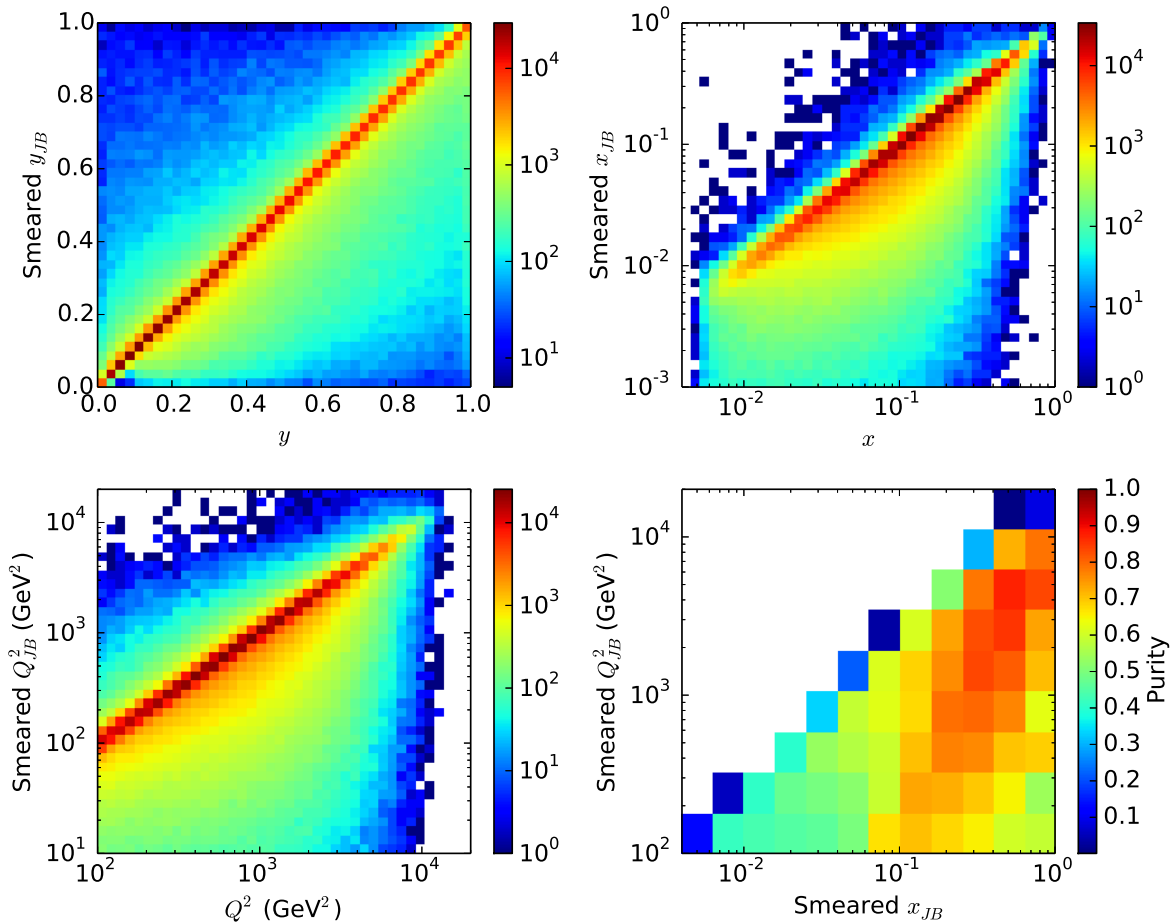,width=0.9\textwidth,angle=0}
\end{center}
\vspace*{-0.6cm}
\caption{\label{fig:jb} [color online] Top left to bottom left panels:
correlations between the reconstructed kinematic variables $y_{JB}$, $x_{JB}$, and $Q^2_{JB}$, 
including detector and radiative effects, and the true generated values. 
Lower right plot: purity in each $x,\,Q^2$ bin, measured as how 
many events that were generated in a bin remain in that bin after reconstruction (see text).
All results are obtained for lepton and proton beam energies of $20\times250\,\mathrm{GeV}$, i.e.
$\sqrt{S}\sim 141\,\mathrm{GeV}$.} 
\end{figure*}
In CC-mediated electron DIS the produced neutrino is not measured,
and the event kinematics have to be reconstructed from the observed hadronic final state.
This is achieved by the JB method \cite{ref:jb} by appropriately summing over all
final-state hadrons $i$ reconstructed within the detector acceptance. This leads to
\begin{align}
\label{eq:kine-jb}
y_{JB} &= \frac{\sum_{i}(E_{i}-p_{z,i})}{2E_{e}}, & Q_{JB}^{2} &= \frac{p_{T,h}^2}{1-y_{JB}},
 & x_{JB} &= \frac{Q_{JB}^{2}}{y_{JB}\,S},
\end{align}
where $E_e$ is the incoming electron beam energy and $p_{T,h}=\left|\sum_{i}\vec{p}_{T,i}\right|$ 
denotes the total transverse momentum of the hadronic final-state particles 
$i$ with measured four-momenta $(E_i,\vec{p}_{T,i},p_{z,i})$. 

Experimental determinations of the kinematic variables (\ref{eq:kine-jb}) are 
not only affected by the radiative corrections calculated with the {\sc Djangoh} generator 
described in Sec.~\ref{sec:django}
but also by the resolution of the detector. To this end, the track properties of the
generated final-state hadrons were smeared according to parametrized resolutions
for particle momenta and energies envisioned for a future EIC detector.  
Momentum resolutions are based on the results of {\sc Geant4} simulations \cite{Agostinelli:2002hh}
of a tracking system comprising TPC, GEM and silicon detectors.
Projected electromagnetic and hadronic calorimeter performances were used to determine
the energy resolutions; for further details on 
studies and plans for an eRHIC detector, see \cite{ref:erhic}.
The obtained momentum resolutions are typically better than a few percent for 
track momenta up to about $60\,\mathrm{GeV}$.
The tracking, electromagnetic and hadronic calorimeter coverage spans
$-3 < \eta < 3$, $-4.5 < \eta < 4.5$, and $2 < \eta < 4.5$ in
pseudo-rapidity, respectively.
In our simulations of CC DIS off polarized neutrons, we assume that
the experiment uses either a polarized $^{2}H$ or a $^{3}He$ beam.
To ensure that the scattering happened on the neutron we require the
spectator protons to be tagged. A commonly used technique would be to
use Roman Pots properly integrated into the interaction region to
guarantee high detection efficiencies of $>98\%$.
Finally, the effects of particle misidentification and finite angular resolution
were assumed to be negligible compared to energy and momentum resolutions.

The results of our studies are summarized in Fig.~\ref{fig:jb} which assumes a
c.m.s.\ energy of $\sqrt{S}\simeq 141\,\mathrm{GeV}$ 
corresponding to eRHIC beam energies of $20\;\mathrm{GeV}\times 250\;\mathrm{GeV}$
and a minimum $Q^2$ of $100\, \mathrm{GeV^2}$.
The JB method generally shows no degradation of the $y$-resolution 
compared to the electron method, see, e.g., Ref.~\cite{Blumlein:2012bf}, 
where the inelasticity is obtained from the scattered electron as
$y_e=1-(1-\cos\theta_e)E^{\prime}_{e}/2E_{e}$.
This is readily understood from the relative $y$-resolutions for both methods: 
$\delta y_{JB}/y_{JB} \sim \mathrm{const}$ and $\delta y_{e}/y_{e} \sim 1/y_e$.
The resolution in $Q^2$ degrades the more of the hadronic transverse momentum $p_{T,h}$
of an event is missed by the detector. Hence, the JB method generally leads to 
poor resolution at low values of $Q^2$ which are, however, of limited interest for CC DIS
measurements. The resolution improves with increasing $Q^2$ as more particles
are scattered into the acceptance of the detector.
As $x_{JB}$ is calculated from $y_{JB}$ and $Q^2_{JB}$, the $x$ resolution generally follows 
that of $Q^2$.
For both $Q^2_{JB}$ and $x_{JB}$, smearing due to detector resolution effects 
and radiative corrections can result in 
reconstructed values significantly deviating from the generated ones.
However, as can be inferred from the first three panels of Fig.~\ref{fig:jb},
at the high $Q^2$ values relevant for CC DIS measurements, the JB method generally
yields good resolutions in all relevant kinematic variables.
We note that for all three kinematic variables detector effects are the
dominant source of smearing.
It was also investigated if any quasi-real photoproduction event could be misreconstructed 
as a high $Q^2$ CC event. The fraction of such events was found to be negligible.

The lower right panel of Fig.~\ref{fig:jb} shows the purity of generated events $N_{gen}$,
defined as $(N_{gen} - N_{out}) / (N_{gen} - N_{out} + N_{in)}$, in
different $x$, $Q^2$ bins for $\sqrt{S}\simeq 141\,\mathrm{GeV}$.
A high purity in a bin indicates that only a 
small fraction of events is smeared in $(N_{in})$
or out $(N_{out})$ of the bin due to detector and radiative effects.
The relatively low purities found at the highest $Q^2$ bin for any given $x$ bin
is caused by binning effects, which can be mitigated by adjusting the
binning in the experiment. In general, radiative effects typically 
deteriorate the purities in each bin by an additional $10\div 20\%$ from
high to low $Q^2$ compared to detector smearing effects.

We note that we obtain quantitatively very similar results also for a lower 
electron beam energy of $10\,\mathrm{GeV}$, i.e., $\sqrt{S}=100\,\mathrm{GeV}$, 
but resulting in a somewhat limited kinematic $x, Q^2$ coverage as will
be illustrated below.

The dominant systematic uncertainty for a measurement of CC DIS at an EIC will be the
error from the measurement of the hadron beam polarization. The
currently best value for polarized protons achieved at RHIC is $3.4\%$. All
other systematic uncertainties for CC DIS will be significantly smaller.

\section{Expectations for spin asymmetries and their impact on
PDF fits\label{sec:impact}}
%
For our detailed phenomenological studies, we will mainly consider a c.m.s.\ energy of
$\sqrt{S}\sim141\,\mathrm{GeV}$ for a future EIC, which offers the largest kinematic
coverage in $x$ and $Q^2$ and hence the best prospects for measurements of CC DIS.
Since this energy is likely to be realized only in a later stage of an EIC, we will also
comment on the feasibility of CC measurements at lower energies such as $\sqrt{S}=100\,\mathrm{GeV}$.
We note again that these two c.m.s.\ energies would correspond to electron beam
energies of 20 and $10\,\mathrm{GeV}$, respectively, for the eRHIC option, which
makes use of the existing $250\,\mathrm{GeV}$ proton beam of RHIC.

%
\begin{figure}[ht]
\begin{center}
\vspace*{-0.6cm}
\epsfig{figure=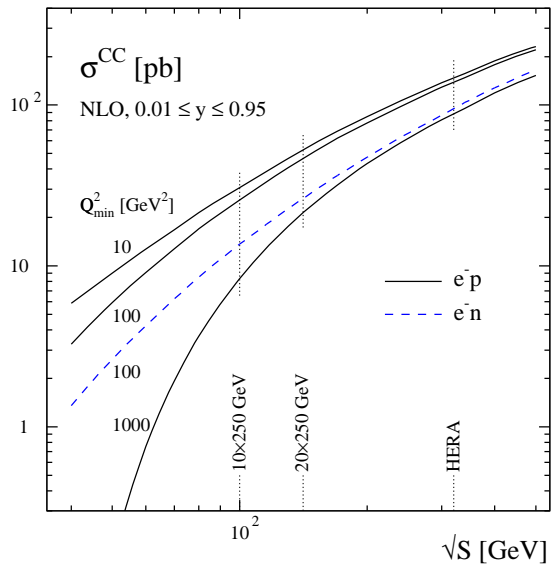,width=0.48\textwidth,angle=0}
\end{center}
\vspace*{-0.5cm}
\caption{\label{fig:sigmatot} Integrated unpolarized CC electron-proton DIS cross section at
NLO accuracy as a function of the c.m.s.\ energy $\sqrt{S}$ for $Q^2>Q^2_{\min}$ and $0.01\le y \le 0.95$.
For comparison we also show the result for electron-neutron scattering (dashed line) for
$Q^2_{\min}=100\,\mathrm{GeV}$.}
\end{figure}
First, we perform some studies of the unpolarized CC cross section to get some idea
about the expected event rate and required integrated luminosity
as a function of $\sqrt{S}$ and the ranges in $x$ and $Q^2$ which are predominantly probed.
In Fig.~\ref{fig:sigmatot} we show the total CC DIS cross section integrated over
the inelasticity $y$ in the range $0.01\le y \le 0.95$ typically accessible in DIS collider experiments.
We present results for both electron-proton and electron-neutron scattering for various
lower bounds on $Q^2$. For all our studies we use the unpolarized NLO MRST parton densities and
corresponding values of the strong coupling \cite{Martin:2001es} as this set was adopted by 
the DSSV collaboration as the reference in evaluating the positivity bound for helicity PDFs 
\cite{deFlorian:2008mr}. 
We note, however, that all our results are only mildly dependent on the choice of unpolarized PDFs.

We recall that both the H1 and the ZEUS collaborations at HERA have successfully performed measurements
of NC and CC DIS off unpolarized protons in a broad range of $x$ and $Q^2$ at a 
c.m.s.\ energy of about $300\,\mathrm{GeV}$ \cite{Aaron:2009aa}.
The observed total CC cross section at HERA is $100\div 200\,\mathrm{pb}^{-1}$ depending on $Q^2_{\min}$.
As can be inferred from Fig.~\ref{fig:sigmatot} the cross section only drops by a factor of about 10 
from HERA to EIC energies, which in terms of expected event rates is more than compensated for 
by the envisioned luminosities of at least a hundred times greater than what was achieved at HERA.

The figure also illustrates the dependence on $Q^2_{\min}$, the lower cut-off on the momentum
transfer squared applied in the calculation of the total CC cross section.
As is expected from the $Q^2$ dependence of the $W^{\pm}$ propagator, see Eq.~(\ref{eq:propagator}),
there is little dependence on $Q^2_{\min}$ as long as $Q^2_{\min} \ll M_W^2$.
However, quite some events would be lost if measurements could be only performed at very
large values of $Q^2\sim M_W^2$. Luckily, as we have demonstrated in Fig.~\ref{fig:jb} in the previous section,
the JB method works sufficiently well for all values of $y$ and 
down to $Q^2\simeq 100\,\mathrm{GeV}^2$ which we will use as a lower cut-off in our studies below. 

%
\begin{figure*}[tbh]
\begin{center}
\vspace*{-0.6cm}
\epsfig{figure=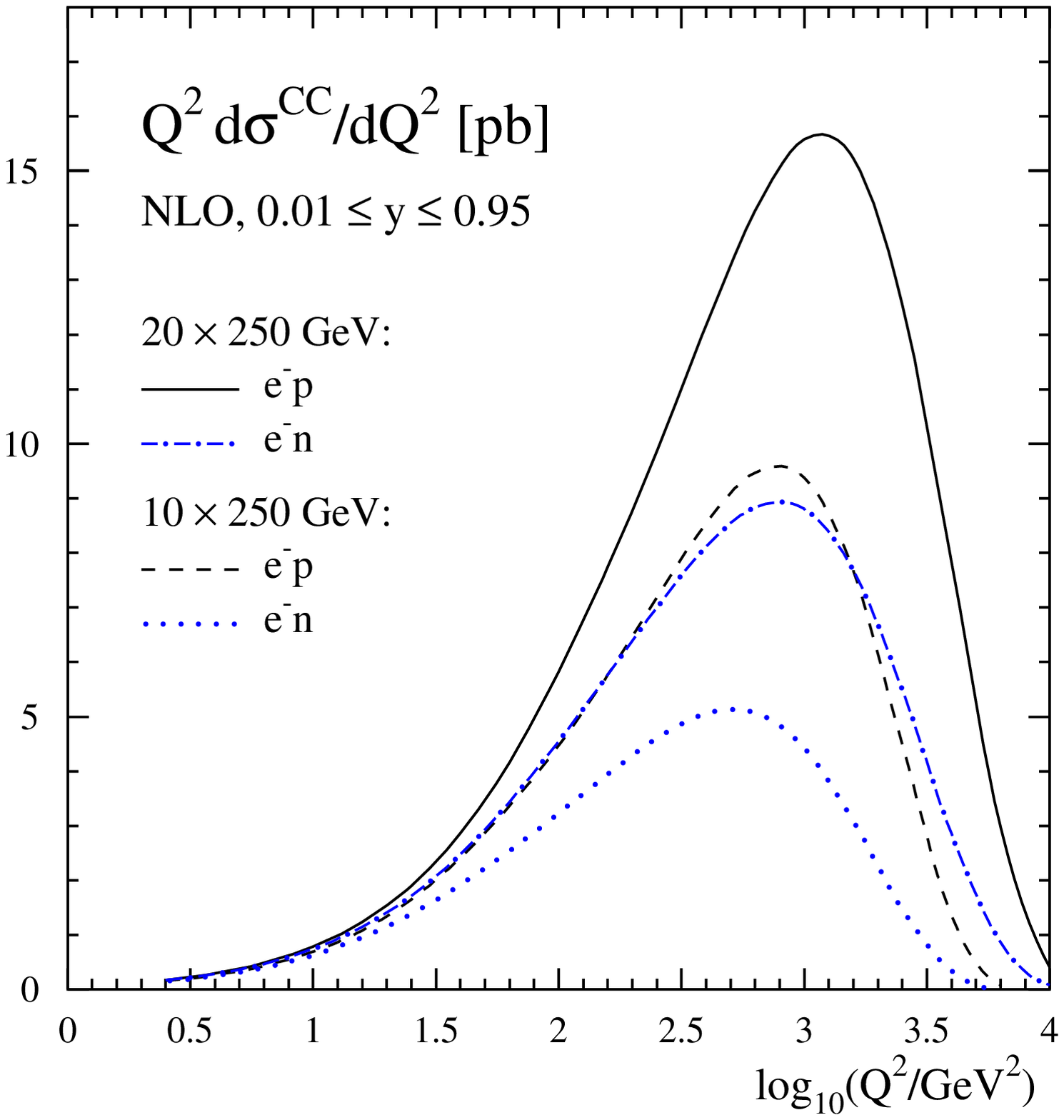,width=0.45\textwidth,angle=0}
\epsfig{figure=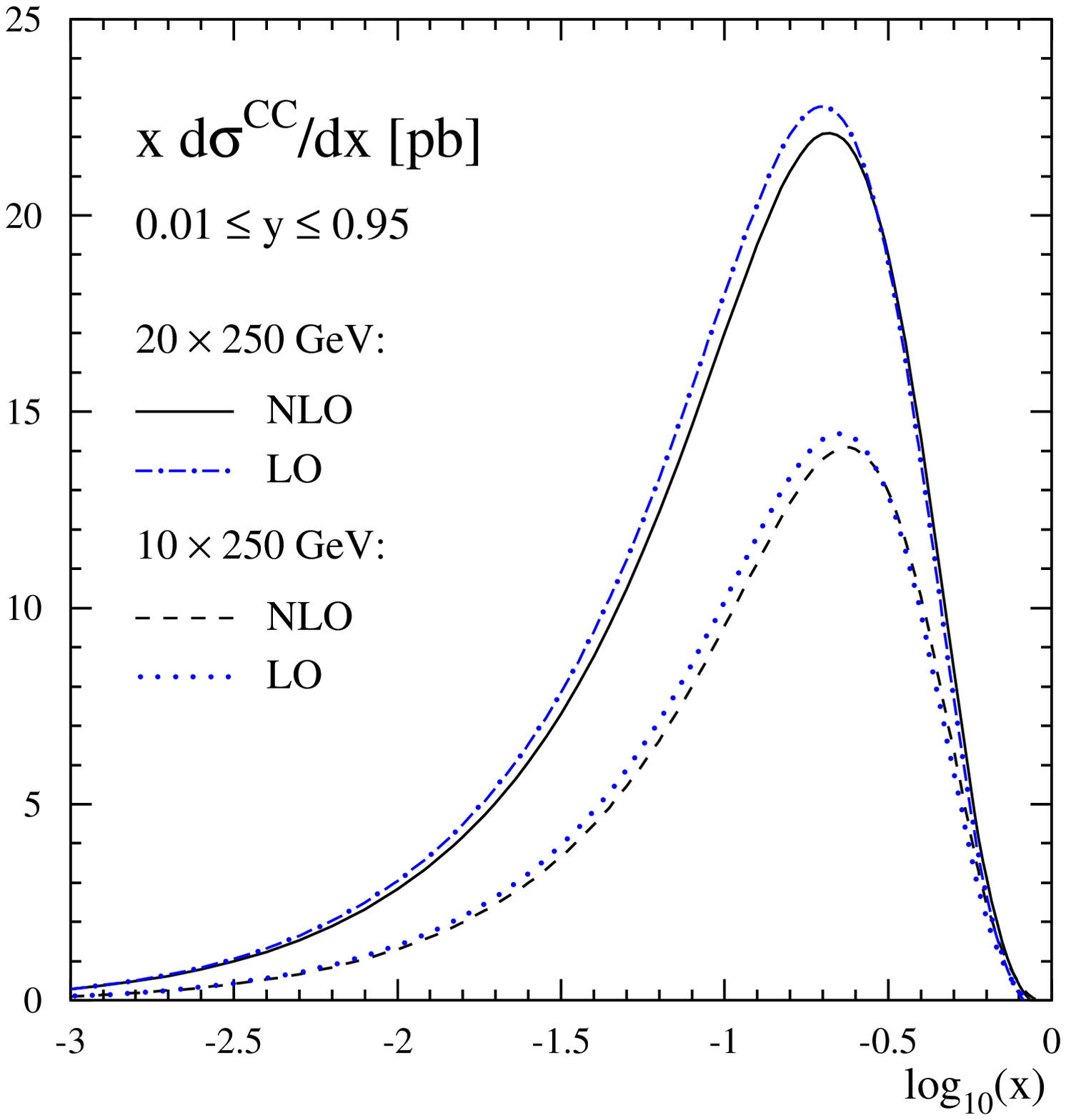,width=0.45\textwidth,angle=0}
\end{center}
\vspace*{-0.5cm}
\caption{\label{fig:unpxsec} {Left panel:} $Q^2$  
differential unpolarized CC cross sections for electron-proton and electron-neutron scattering at NLO accuracy for
two different c.m.s.\ energies and integrated in the range $0.01\le y \le 0.95$.
{Right panel:} similar, but now showing the LO and NLO $x$ distributions for $e^-p$ CC DIS.}
\end{figure*}
Finally, Fig.~\ref{fig:sigmatot} also shows the difference between CC DIS off a proton and a neutron
target. The $e^-n$ CC cross section is about a factor of two smaller than the one for $e^-p$ scattering
which is readily understood from the applied isospin rotation and the fact that $u(x)>d(x)$. 
Nevertheless, given the expected integrated luminosities at an EIC, measurements with an effective
neutron beam should be equally feasible as those performed in $e^{-}p$ CC DIS.

Figure~\ref{fig:unpxsec} sheds more light on how CC DIS events will be distributed in $Q^2$
(left panel) and $x$ (right panel) at the two considered EIC energies.
Presented are the single differential distributions in $\log (Q^2)$ and $\log (x)$ for $e^-p$ scattering, 
as above integrated in the range $0.01\le y \le 0.95$. In case of the $Q^2$ distribution we
compare with CC DIS on an effective neutron beam,  while for the $x$ differential cross section we
study the effect of NLO corrections.
As can be seen from the left panel, the bulk of events are centered at $Q^2\simeq 1000\,\mathrm{GeV}$ 
but with a fairly broad
tail down to lower values of $Q^2$, which explains the differences in the results
obtained for $Q^2_{\min}=100$ and $1000\,\mathrm{GeV}^2$ shown in Fig.~\ref{fig:sigmatot}. 
The peak of the $Q^2$ distribution, and hence the number of expected events, 
drops by almost a factor of two when lowering the c.m.s.\ energy to $100\,\mathrm{GeV}$ or, similarly,
upon replacing the proton by an effective neutron beam.
Due to the kinematic relation $x=Q^2/(yS)$, the $x$ differential distribution peaks at values
of about $x\sim 0.2$ but again with significant tails. 
Since we can safely assume an integrated luminosity of $10\,\mathrm{fb}^{-1}$ \cite{Accardi:2012hwp}, CC DIS measurements
at an EIC will be feasible down to $x$ values close to the kinematic limit for $Q^2_{min}=100\,\mathrm{GeV}$.
As we have already mentioned in Sec.~\ref{sec:theory}, the NLO corrections are rather modest in the
entire kinematic range relevant for CC DIS measurements.

%
\begin{figure}[h]
\begin{center}
\vspace*{-0.6cm}
\epsfig{figure=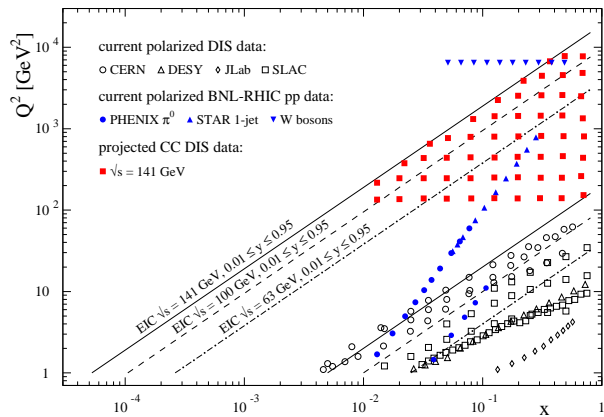,width=0.48\textwidth,angle=0}
\end{center}
\vspace*{-0.5cm}
\caption{\label{fig:kine} [color online] Kinematic range in $x$ and $Q^2$
accessible with three different c.m.s.\ energies at an EIC for $0.01\le y \le 0.95$.
The solid squares indicate the projected CC DIS data for $\sqrt{S}\sim 141\,\mathrm{GeV}$. 
The other symbols illustrate the coverage of the currently available suite of measurements 
from spin-dependent DIS and $pp$ experiments.}
\end{figure}
A reasonable binning in $x$ and $Q^2$ should allow for detailed studies of PDFs in the large $Q^2$ region.
To quantify the impact of future CC DIS measurements at an EIC with polarized protons and neutrons on our
understanding of helicity PDFs, we assume the binning already used in the lower right panel of Fig.~\ref{fig:jb}
[and for the estimates of radiative corrections shown in Figs.~\ref{fig:rc1} and \ref{fig:rc2}].
It can be further optimized, of course, once actual data become available. For $x>0.1$ the estimated
purities are in general above $70\%$, while they drop into the $40$-$50\%$ range for $0.01\le x \le 0.1$.
The error bars on the generated cross sections and single-spin asymmetries properly reflect the kinematic smearing 
due to radiative corrections and detector effects, i.e., corrections from the unfolding of the ``true'' kinematic
variables.
  
Figure~\ref{fig:kine} illustrates the coverage of the simulated CC DIS data with $Q_{\min}^2=100\,\mathrm{GeV}^2$
in the $x$ and $Q^2$ plane, where we also overlay existing DIS data from fixed-target experiments. 
For the $pp$ inclusive jet and pion data from RHIC \cite{Aschenauer:2013woa}
we have chosen $x=2p_T/\sqrt{S}$ to give a rough idea
of the lowest $x$ values probed for measurements at a given $p_T$. The actual $x$ range is, of course, very
broad due to the complicated convolutions of the hard scattering cross sections with the two PDFs needed 
to describe $pp$ collisions.
We note that only existing and upcoming $W$ boson production data from RHIC \cite{Aschenauer:2013woa}
can access $x$ and $Q^2$ values comparable to
the range covered by CC DIS at an EIC. Projections of their expected impact can be found in Ref.~\cite{Aschenauer:2013woa}.

The lines in Fig.~\ref{fig:kine} indicate the allowed kinematic range for $0.01\le y \le 0.95$ for three different conceivable
c.m.s. energies for an EIC. The upper line refers to $y=0.95$, while the lower one corresponds to $y=0.01$.
The highest $\sqrt{S}$, for which we have generated the projected CC DIS data, offers,
of course, the best coverage. At $x>0.1$ one has access to almost two orders of magnitude in $Q^2$ for any
given fixed $x$ value, which allows one to study QCD evolution effects.
However, with $\sqrt{S}=100\,\mathrm{GeV}$, which is expected to be initially available at eRHIC \cite{ref:erhic},  
a similar type of measurements is still feasible but with lower rates in each bin, cf.\ Fig.~\ref{fig:unpxsec}.
Even lower c.m.s.\ energies cut further into the accessible $x$ and $Q^2$ range at further reduced
event rates and, hence, do not seem to be very attractive anymore. For instance, at $\sqrt{S}\simeq 63\,\mathrm{GeV}$,
corresponding to $10\times 100\,\mathrm{GeV}$ collisions at eRHIC,
only about 3 bins will remain above $Q^2=1000\,\mathrm{GeV}^2$.

%
\begin{figure}[h]
\begin{center}
\vspace*{-0.6cm}
\epsfig{figure=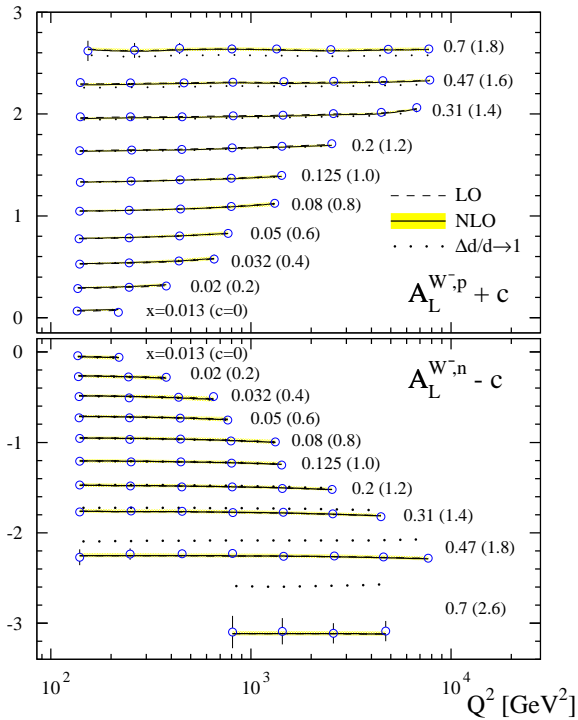,width=0.48\textwidth,angle=0}
\end{center}
\vspace*{-0.5cm}
\caption{\label{fig:asym} [color online] Projected single-spin asymmetries $A_L^{W^{-},p}$ ({top panel}) and
$A_L^{W^{-},n}$ ({bottom panel}) for $\sqrt{S}\sim 141\,\mathrm{GeV}$ compared to LO and NLO 
calculations using the DSSV helicity densities. The dotted line shows an alternative DSSV set 
which enforces $\Delta d/d\to 1$ as $x\to 1$ (see text).
The shaded bands correspond to the $\Delta \chi^2=8$ uncertainty estimates for
the DSSV PDFs. Note that a constant $c$ is added to each bin as indicated.}
\end{figure}
With all these preparatory studies at hand, 
we turn now to our main results: the projected CC single-spin asymmetries and their potential impact in a global
QCD analysis of helicity PDFs.
In Fig.~\ref{fig:asym} we show the simulated spin asymmetries for CC DIS off polarized proton and
neutron beams assuming a c.m.s.\ energy of $\sqrt{S}=141\,\mathrm{GeV}$.
Note that we have added a constant $c$ to the asymmetry in each $x$ bin in order to clearly separate the results
obtained in different bins.
The top panel shows $A_L^{W^{-},p}$, which is positive 
and without the rescaling constant takes values ranging from a few percent at the smallest $x$ 
value to more than $80\%$ at $x\simeq 0.7$. 
$A_L^{W^{-},n}$ (bottom panel) is negative and somewhat smaller in size, reaching about
$-50\%$ at $x\simeq 0.7$. 
As mentioned above, the estimated errors reflect the statistical accuracy for an integrated luminosity
of $10\,\mathrm{fb}^{-1}$ after unfolding detector smearing and radiative effects.

The behavior of the asymmetries in Fig.~\ref{fig:asym} is best understood at LO accuracy, 
which is a very good approximation as can also be inferred from Fig.~\ref{fig:asym}.
At LO, the asymmetry (\ref{eq:asymdef}) simplifies to (recalling that $g_L=0$,
$g_5\to -F_1$ and $2g_1\to F_3$):
\begin{equation}
\label{eq:asymlo}
A^{W^-,p} = \frac{2b g_1^{W^-,p}-a g_5^{W^-,p}}
{a F_1^{W^-}+b F_3^{W^-}} \; ,
\end{equation}
where $a=2(y^2-2y+2)$, $b=y (2-y)$, and similarly for $A_L^{W^{-},n}$. 
To proceed, we approximate (\ref{eq:asymlo}) further by studying some limiting cases for $y$,
assuming that contributions from strange and charm quarks are negligible. 
The results are summarized in Tab.~\ref{tab:asym}.
%
\begin{table}[th!]
\caption{\label{tab:asym} Approximate behavior of the LO single spin asymmetry (\ref{eq:asymlo}) for
$e^-p$ and $e^-n$ CC DIS for certain fixed values of $y$.}
\begin{ruledtabular}
\begin{tabular}{ccccc}
                  & $y\rightarrow 0$       & $y=1/2$           & $y\rightarrow 1$\\ \hline \\
$A_L^{W^-,p}$     & $\frac{\Delta u(x) - \Delta \bar{d}(x)}{u(x)+\bar{d}(x)}$  & 
                    $\frac{4\Delta u(x) -\Delta \bar{d}(x)}{4 u(x) + \bar{d}(x)}$ &  
                    $\frac{\Delta u(x)}{u(x)}$ \\ [3mm]
$A_L^{W^-,n}$     & $\frac{\Delta d(x) - \Delta \bar{u}(x)}{d(x)+\bar{u}(x)}$  & 
  	                $\frac{4\Delta d(x) -\Delta \bar{u}(x)}{4d(x) + \bar{u}(x)}$& 
	                $\frac{\Delta d(x)}{d(x)}$	      \\            \\
\end{tabular}
\end{ruledtabular}
\end{table}
%

%
\begin{figure}[th]
\begin{center}
\vspace*{-0.6cm}
\epsfig{figure=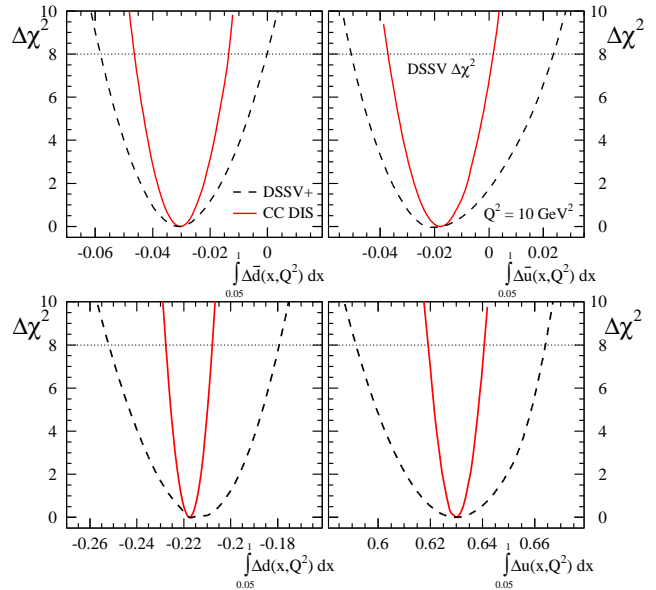,width=0.48\textwidth,angle=0}
\end{center}
\vspace*{-0.5cm}
\caption{\label{fig:profiles} [color online] $\chi^2$ profiles for the first moments of the
helicity PDFs $\Delta u$, $\Delta d$, $\Delta \bar{u}$, and $\Delta \bar{d}$ truncated
to the region $0.05\le x \le 1$ at $Q^2=10\,\mathrm{GeV}^2$ and evaluated with (solid lines)
and without (dashed lines) the projected CC DIS data shown in Fig.~\ref{fig:asym}.
The dotted line indicates the $\Delta \chi^2$ tolerated in the original DSSV analysis \cite{deFlorian:2008mr}.}
\end{figure}
For large values of $x$ and at all $y$, $e^-p$ and $e^-n$ CC DIS data essentially track the polarization values $\Delta q/q$ 
for $u$ and $d$ quarks, respectively. In the DSSV fit they approach
$1$ for $\Delta u/u$ and approximately $-0.6$ for $\Delta d/d$; see Fig.~5 in the second reference of \cite{deFlorian:2008mr}.
At smaller values of $x$, where valence quark contributions are small and $\Delta q\simeq \Delta \bar{q}$, various combinations
of light sea quark polarizations can be studied depending on $y$; see Tab.~\ref{tab:asym}. 
Only at large $y$, data are essentially sensitive
to $\Delta \bar{u}/\bar{u}$ and $\Delta \bar{d}/\bar{d}$ for $e^-p$ and $e^-n$ scattering, respectively.

While $\Delta u/u$ at large $x$ is pretty well constrained from existing fixed-target DIS data, there are 
theoretical expectations based on ``helicity retention'' \cite{Avakian:2007xa}
that $\Delta d/d$ should saturate at 1 as $x\to 1$. Such a behavior would require a dramatic change in 
the trend seen with present data \cite{deFlorian:2008mr}, 
which constrain $\Delta d/d$ to negative values around $-0.5$ up to $x \simeq 0.6$ for the modest $Q^2$ values accessible
in fixed-target experiments. 
From the considerations above, measurements of $A_L^{W^-,n}$ would be particularly suited to study a possible 
sign change in $\Delta {d}/{d}$ at large values of $x$. 

To make this more quantitative, the dotted lines in Fig.~\ref{fig:asym}
are obtained with a special set from DSSV \cite{deFlorian:2008mr} where $\Delta d/d \to 1$ is enforced by
adding extra terms to the functional form assumed in their analysis, leading to 
a sign change at $x\simeq 0.67$ at a scale of $Q^2=1\,\mathrm{GeV}^2$. 
Note that such a behavior is way outside the DSSV uncertainty estimates ($\Delta \chi^2=8$) based on 
their standard functional form \cite{deFlorian:2008mr},
which are indicated as shaded bands in each $x$ bin in Fig.~\ref{fig:asym}.

Since CC DIS probes the PDFs at rather large scales, QCD evolution shifts the assumed node
in $\Delta d/d$ to smaller $x$ values, such that it significantly impacts theoretical
expectations for $A_L^{W^-,n}$ already at $x\simeq 0.4$. As expected, $A_L^{W^-,p}$
changes only very little, mainly due to minor adjustments in $\Delta u$ and the sea quark densities 
in the special DSSV set in order to respect sum rules.
We note that $W^-$ boson production at large forward rapidities in polarized $pp$ collisions at
RHIC is in principle also sensitive to $\Delta d/d$ at large $x$; see Fig.~15 in the second
reference of \cite{deFlorian:2008mr}. 

Finally, we estimate the impact of the projected data shown in Fig.~\ref{fig:asym} in a
global QCD analysis of helicity PDFs. To this end, we closely follow the methodology of the
DSSV collaboration \cite{deFlorian:2008mr} and the procedures outlined in a similar study 
based on simulated EIC data for polarized NC (photon-exchange) DIS data in Ref.~\cite{Aschenauer:2012ve}. 
To visualize the expected improvements in $\Delta u$, $\Delta d$, $\Delta \bar{u}$, and $\Delta \bar{d}$ 
due to CC DIS data we study their truncated moments, defined as
\begin{equation}
\label{eq:truncmom}
\Delta q(x_{\min},x_{\max},Q^2) = \int_{x_{\min}}^{x_{\max}} \Delta q(x,Q^2)\,dx\;.
\end{equation}
For each parton flavor $f$ we minimize $\chi^2$ with an additional constraint on the value
of its truncated moment (\ref{eq:truncmom}), implemented through a Lagrange multiplier,
to map out the corresponding $\chi^2$ profile away from its best fit value.
The width of the profiles, read off at a certain increase in $\Delta \chi^2$ that is still
tolerated for a good fit, serves as an estimator for the uncertainty on the truncated moment
$\Delta q(x_{\min},x_{\max},Q^2)$.

As reference profiles we take the full dataset used in the DSSV global analysis \cite{deFlorian:2008mr}
augmented by the latest DIS data from the COMPASS fixed-target experiment \cite{Alekseev:2009ac}. 
This fit is often referred to as DSSV+ \cite{deFlorian:2011ia,Aschenauer:2013woa}.
Figure~\ref{fig:profiles} shows the profiles for $\Delta q(x_{\min},x_{\max},Q^2)$, where
$q=u,d,\bar{u},$ and $\bar{d}$, for $x_{\min}=0.05$ and $x_{\max}=1$ at $Q^2=10\,\mathrm{GeV}^2$. 
The dashed lines refer to the DSSV+ fit, and the solid lines include the projected CC DIS data
shown in Fig.~\ref{fig:asym}
in the global analysis procedure. The value for $x_{min}$ takes into account that the 
constraints from CC DIS data on PDFs obtained at some large value of $Q^2$ translate into constraints 
at larger values of $x$ at smaller values of $Q^2$ due to QCD evolution effects. Since our
projected data cover $x$ values down to $x\simeq 0.01$ at $Q^2 \simeq 100\,\mathrm{GeV}^2$,
see Fig.~\ref{fig:kine}, $x_{\min}=0.05$ appears to be an appropriate choice. 

Clearly, CC DIS data will greatly help to further our understanding of 
helicity PDFs for $u$ and $d$ quarks and antiquarks. Note that the sensitivity of $A_L^{W^-,n}$ to 
a possible $\Delta d/d\to 1$ for $x\to 1$ as discussed above is not even included in the
estimates shown in Fig.~\ref{fig:profiles} as they are based on the standard functional
form of DSSV.
It should be also stressed that CC DIS data will lead to constraints on individual quark
flavors independent of other non-perturbative inputs such as fragmentation functions (FFs).
Together with $W$ boson asymmetries from RHIC they offer invaluable checks on
current estimates of flavor separated helicity PDFs \cite{deFlorian:2008mr,Leader:2010rb} 
which are solely based on 
semi-inclusive DIS data taken at relatively low values of $Q^2$ where, 
in addition to uncertainties from FFs \cite{Epele:2012vg}, 
power or higher-twist corrections can still be of some importance.

\section{Summary and Outlook}
%
We have presented a detailed phenomenological study of CC-mediated DIS off longitudinally 
polarized protons and neutrons at a future Electron-Ion Collider. 
Our simulations have been based on an updated version of the event generator package
{\sc Djangoh}, which previously could only handle unpolarized nucleons in DIS.

We have carefully explored the applicability of the Jacquet-Blondel method for reconstructing
the relevant kinematic variables from the observed hadronic final state.
Detector effects were found to be the dominant source of kinematic smearing
while the radiative corrections included in {\sc Djangoh} typically add an additional $10$-$20 \%$. 
Error estimates for our simulated charged current single-spin asymmetries in electron-proton and electron-neutron
scattering properly include unfolding corrections from imperfect purities in each bin.

To estimate the impact of the generated charged current single-spin asymmetries on
determinations of helicity parton distributions, we have performed global QCD analyses at next-to-leading order accuracy
including studies of uncertainties based on the Lagrange multiplier technique.
It was demonstrated that charged current DIS measurement at an EIC would be a very valuable addition to
the suite of existing data, as they provide independent constraints on the $u$ and $d$ quark and antiquark
helicity densities, free of ambiguities from hadronization.
In addition, measurements of charged current DIS off a longitudinally polarized effective
neutron target, with spectator protons being tagged, have been shown to 
be particularly sensitive to theoretical expectations based on helicity retention arguments that the 
$d$ quark polarization should saturate and align with the nucleon's spin direction at large momentum
fractions $x$.
All our studies were performed assuming a c.m.s.\ energy of $\sqrt{S}\simeq 141\,\mathrm{GeV}$,
however, energies down to about $100\, \mathrm{GeV}$, which are expected to be available at an
initial version of an EIC, appear to be sufficient as well. In each case, only modest 
integrated luminosities of about $10\,\mathrm{fb}^{-1}$ are needed to perform inclusive charged current DIS measurements
at an EIC.

We note that there are various other avenues for electroweak measurements at an EIC that can be pursued
based on the tools developed and studies performed in this paper. A natural extension would be to
look into charged current mediated semi-inclusive DIS. Identified kaons may provide some sensitivity
to the strangeness polarization at medium-to-large values of the momentum fraction $x$ and large $Q^2$.
The relevant next-to-leading corrections were computed recently in \cite{deFlorian:2012wk}.
Another way to determine strange sea distributions at an EIC could be provided by measurements of 
charged current mediated charm production in DIS. Next-to-leading order QCD calculations for this process
have already been performed quite some time ago \cite{Kretzer:1999nn} when the prospect of
operating HERA with polarized hadron beams was discussed.
Both types of processes would require further additions to the {\sc Djangoh} event generator before
detailed phenomenological studies, including detector effects and radiative corrections, can be pursued.
Lastly, it might be interesting to determine whether or not an EIC could provide determination of
electroweak vector and axial-vector couplings with an accuracy better than what was already
achieved at HERA. 
This would require detailed studies of neutral current electroweak effects, in particular, from
photon-$Z$ boson interference contributions.

\section*{Acknowledgments}
%
We are grateful to W.\ Vogelsang for useful discussions about the results shown in the
EIC White Paper.
We acknowledge support by the U.S.\ Department of Energy under contract number DE-AC02-98CH10886
and by a "Laboratory Research and Development" grant (LDRD 12-034) from Brookhaven National Laboratory.
H.S.\ has been supported in part by the DFG in the SFB 1044.

\end{document}